\documentclass{svjour3}                     % onecolumn (standard format)
\smartqed  % flush right qed marks, e.g. at end of proof
\usepackage{graphicx}
\usepackage{natbib}
\bibpunct{(}{)}{;}{a}{}{,}
\usepackage{mathptmx}      % use Times fonts if available on your TeX system
\newcommand{\aap}{Astron. Astrophys.\ }
\newcommand{\bac}{Bull. Astron. Instit. Czechoslovakia\ }
\newcommand{\emp}{Earth Moon Planets\ }
\newcommand{\grl}{Geophys. Res. Lett.\ }
\newcommand{\jrasc}{J. R. Astron. Soc. Canada\ }
\newcommand{\maps}{Meteorit. Planet. Sci.\ }
\newcommand{\mnras}{Mon. Not. R. Astron. Soc.\ }
\newcommand{\wgn}{WGN, Journal of the IMO\ }

\journalname{Earth Moon Planets}
\begin{document}

\title{Spectral, photometric, and dynamic analysis of eight Draconid meteors%\thanks{Grants or other notes
}


\author{Ji\v r\'\i\ Borovi\v cka \and Pavel Koten \and Luk\'a\v s Shrben\'y \and Rostislav \v Stork \and Kamil Hornoch}

\authorrunning{J. Borovi\v cka et al.} % if too long for running head

\institute{Ji\v r\'\i\ Borovi\v cka \and Pavel Koten \and Luk\'a\v s Shrben\'y \and Rostislav \v Stork \and Kamil Hornoch \at
              Astronomical Institute, Academy of Sciences, 251 65 Ond\v rejov Observatory, 
Czech Republic \\
                \email{jiri.borovicka@asu.cas.cz}          
}

\date{Received: date / Accepted: date}

\maketitle

\begin{abstract}
We analyzed spectra, trajectories, orbits, light curves, and decelerations 
of eight Draconid meteors observed from Northern Italy 
on October 8, 2011. Meteor morphologies of two of the
meteors are also presented, one of them obtained with a high resolution camera.
Meteor radiants agree with theoretical predictions, with a hint that some meteors may belong 
to the pre-1900 meteoroid trails.
The spectra confirm that Draconids have normal chondritic composition of main elements (Mg, Fe, Na). There are,
nevertheless, differences in the temporal evolution of Na line emission. The differences are correlated with
the shapes of the light curves and the deceleration rates. Our data confirm that Draconids are porous
conglomerates of grains, nevertheless, significant differences in the atmospheric fragmentation of cm-sized 
Draconids were found. Various textures with various resistance to fragmentation exist among 
Draconid meteoroids and even within single meteoroids.
\end{abstract}

\section{Introduction}
\label{intro}
The Draconid meteor shower, almost unnoticeable in most years, produced intense meteor storms in 1933 and 1946
as well as outbursts in 1985 and 1998 \citep{JennBook} and moderate activity in some other years, 
e.g.\ 2005 \citep{Brown05, Koten05}. The parent body of the shower is the Jupiter family comet 21P/Giacobini-Zinner. 
The mechanism of the outbursts is now well understood, which allowed the prediction of another outburst 
on October 8, 2011 \citep{JennBook, Jeremy1, Maslov}. The outburst was successfully observed by a number of teams 
using various techniques \citep[e.g.][]{Kero, Toth, Trigo, Madiedo, Jeremy2}. Here we report the results of our
simultaneous spectral, photometric, and  dynamic observations of eight relatively bright Draconid meteors. We
concentrate on determination of physical and chemical properties of the meteoroids. This is a continuation of our 
previous work \citep{Bor05}, where we analyzed one bright and six faint (with no spectral data) Draconids observed
in 2005.

\section{Instrumentation}
\label{obs}
Our observations were performed at two sites in Northern Italy, 
Brenna (site A, longitude 9.18795$^\circ$, latitude 45.73371$^\circ$, altitude 333 m)
and Barengo (site B,  8.50505$^\circ$, 45.56606$^\circ$, 238 m). 
Both sites were equipped with high frame rate image intensified video cameras MAIA
\citep{MAIA} for measuring meteor trajectories, velocities and light curves. The trajectory 
work was supported by one supplementary video on site A and one DSLR camera at each site.
For studying meteor morphologies, a  high-resolution (HDV format) non-intensified video camera was used
for the first time. 
The shutter speed was set to 1/120 s, so that quasi-instantaneous images 
of the meteor could be taken.
The camera had low sensitivity and small field of view, so the chances of
capturing a meteor were not high. Nevertheless, one meteor was captured and its shape 
could be studied in much higher detail than with other cameras used.
 
The spectra were taken with our image intensified spectral video camera. We used longer focal
length than usual to increase spectral resolution, at the cost of a smaller field of view.
More details of our expedition to Italy and the observation strategy are given in our first report
\citep{IMC}. Technical details of the cameras are given in Table~\ref{instrum}.

\begin{table}
\caption{Instruments used in this work.}
\label{instrum}
\begin{tabular}{llllllll}
\hline\noalign{\smallskip}
&  MAIA & Suppl. & Spectral & HDV  & All-sky & Wide-field \\
& video & video & video & video & photo & photo \\
\noalign{\smallskip}\hline\noalign{\smallskip}
Camera & GigE & DFK31 & Panasonic  & Canon  & Canon EOS  & Canon EOS  \\
 &Vision  &  & NVS88 &Legria HV40 & 5D Mark II & 450D \\[0.5ex]
Image &Mullard & Dedal 41 &Mullard & -- & -- & -- \\
intensifier &  XX1332 &  &  XX1332  \\[0.5ex]
Grating & -- & --  & 600 groo- & -- & -- &  -- \\
&&& ves/mm \\[0.5ex]
Lens & 1.4/50 & 1.4/50 & 2/85 & native & 2.8/15 & 3.5/10 \\[0.5ex]
Resolution & 776$\times$582 & 1024$\times$768 & 768$\times$576 & 1920$\times$1080 &
5616$\times$3744 & 4272$\times$2848 \\[0.5ex]
Frame rate & 61.15 & 15 & 25 & 25& -- & -- \\[0.5ex]
Exposure & 0.01635 & 0.0667 & 0.04 & 0.008 & 30 & 20 \\
time, s \\[0.5ex]
ISO &--&--&--&--& 1600 & 1600 \\[0.5ex]
Field of &  $50^\circ$ & $30^\circ$ & $30^\circ$ & 18$\times$10$^\circ$ & $180^\circ$ & 95$\times$70$^\circ$ \\
view &&&&&diagonal \\[0.5ex]
Pixel scale$^a$ & 0.106 & 0.044 & 0.048 & 0.0095 & 0.024 & 0.029 \\
deg/pixel & \\[0.5ex]
Limiting &+5&+4&+2&0& 0 & +1 \\
magnitude \\[0.5ex]
Recording & PC & laptop & S-VHS & MiniDV & card & card \\
&&&tape& tape \\[0.5ex]
Site & A,B & A & B & B & B & A \\[0.5ex]
Captured & 4,5,8 (A) & 2--8& 1--8 & 6 & 2,3,4,6 & 1,2,4--8\\
meteors & 1--8 (B) \\
\noalign{\smallskip}\hline
\end{tabular} \\[0.5ex]
$^a$in the center of the field
\end{table}

\section{Analyzed meteors}
\label{basicdata}

\begin{figure}
  \includegraphics[width=0.85\linewidth]{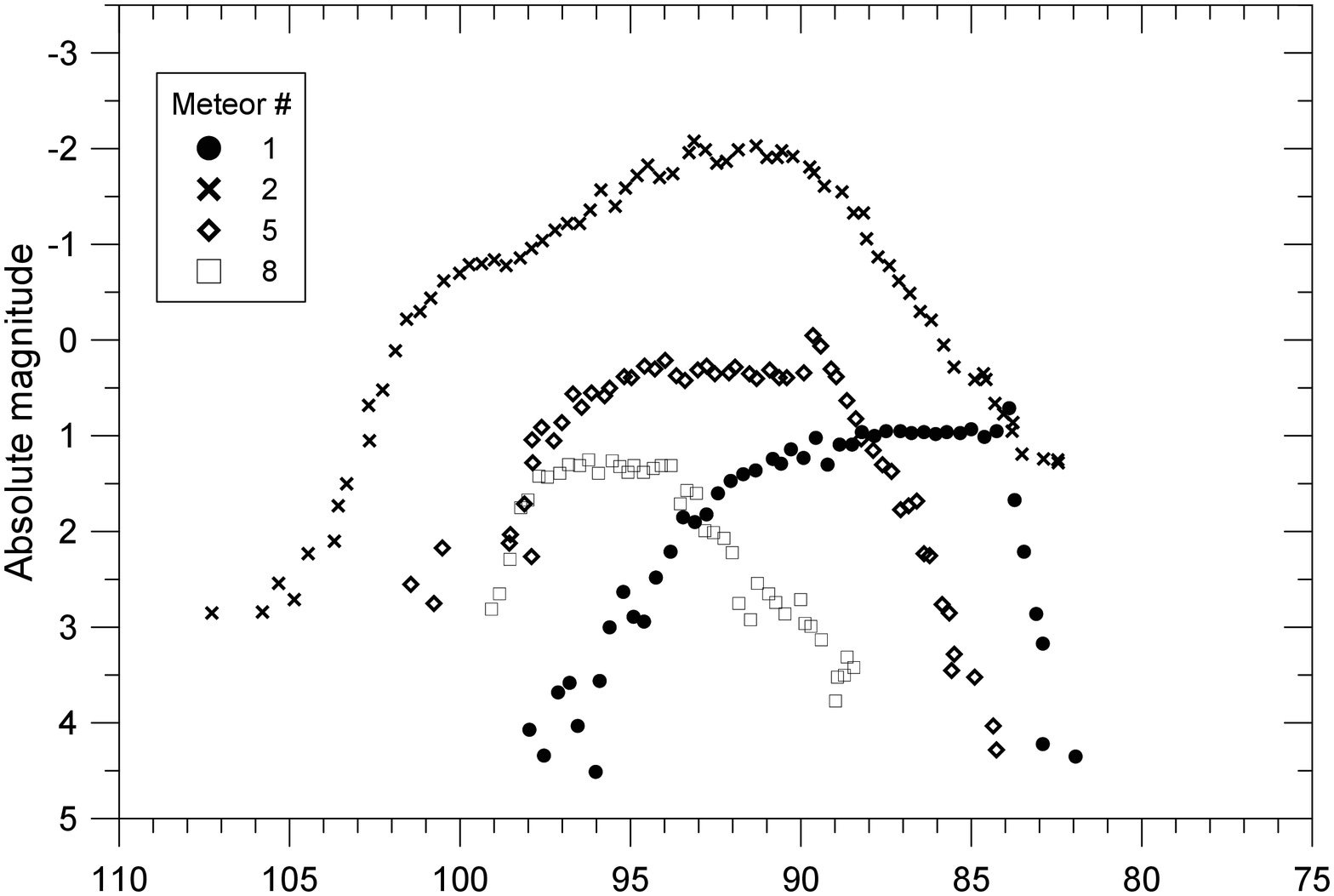}
  \includegraphics[width=0.85\linewidth]{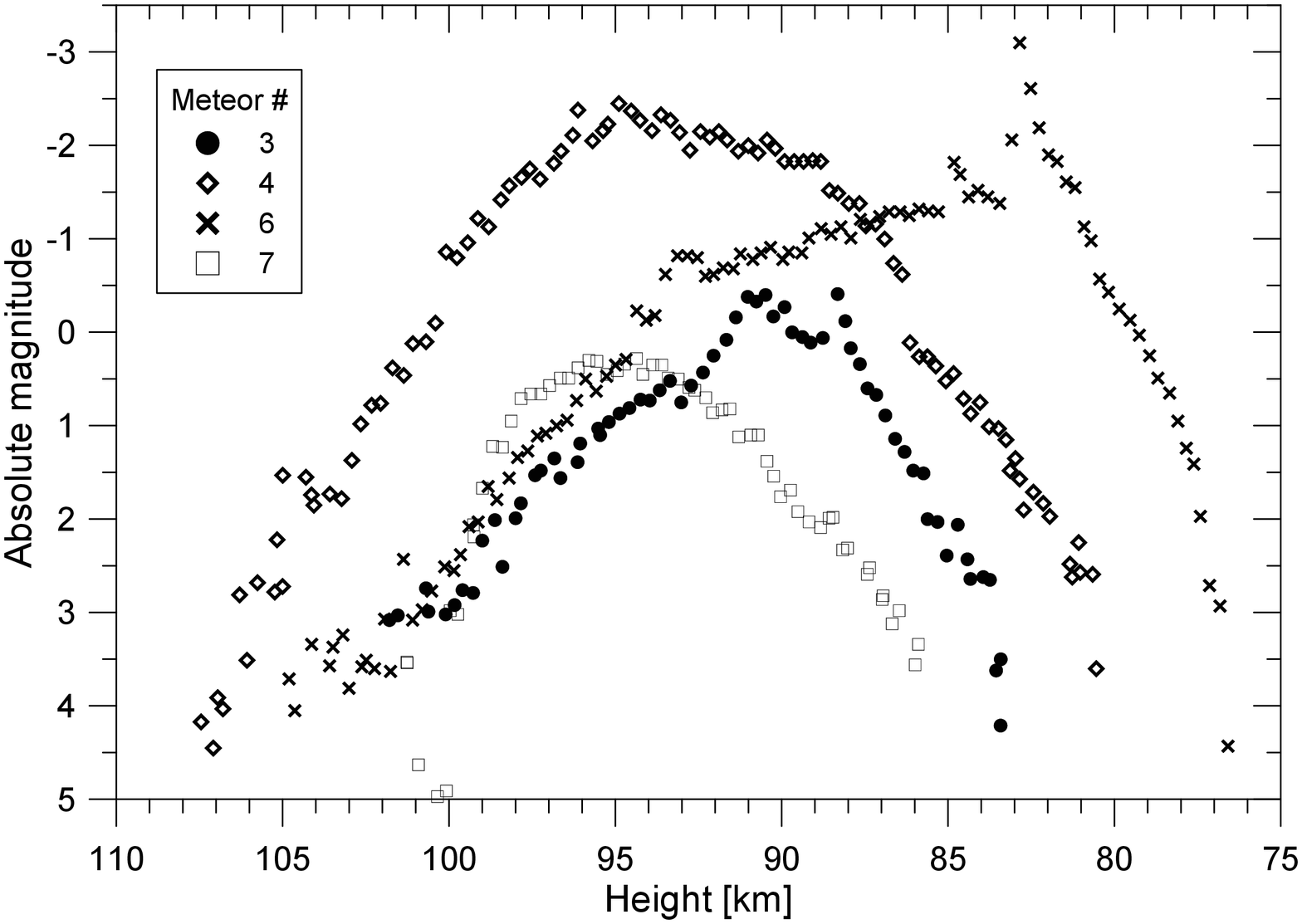}
\caption{Meteor light curves as observed by the MAIA video cameras. The curves are presented in two plots
to improve their visibility.}
\label{LC}      
\end{figure}

Eight Draconid meteor spectra were captured. All eight meteors were also observed by imaging
cameras at both stations, so their trajectories could be computed. 
All meteor records listed in Table~\ref{instrum}, except those from spectral and HDV cameras, were 
combined to compute the trajectories.
Table~\ref{meteors} gives the basic
data on the meteors, including their atmospheric trajectories. The time of appearance,
photometric mass, zenith distance of the radiant,  $z_{\rm R}$, height of first registration,  $h_{\rm beg}$,
height at maximum brightness, $h_{\rm max}$, the absolute maximum brightness, $M_{\rm max}$,
height of disappearance, $h_{\rm end}$, range to the camera, which was used to measure 
the light curve, and
the convergence angle, $Q$, between the planes as seen from both sites are given. The photometric
mass was computed using the luminous efficiency of \citet{tau}. For meteors that exhibited flares,
the height of maximum is given to 0.1 km. Conversely, the light curves of some meteors were flat near the maxima.
In those cases a
height range is given for the maximum. Meteor 3 had two maxima of equal brightness. The actual light
curves are plotted in Fig.~\ref{LC}. The plane convergence
angle was quite low for meteors 5 and 8, nevertheless, we were able to obtain relatively good trajectory solutions.

\begin{table}
\caption{List of studied meteors.}
\label{meteors}
\begin{tabular}{llllllllll}
\hline\noalign{\smallskip}
\#&	Time 	&Mass & $z_{\rm R}$ & $h_{\rm beg}$ & $h_{\rm max}$ & $M_{\rm max}$ & $h_{\rm end}$ & Range & $Q$ \\
&UT&g&$^\circ$& km & km & mag & km & km  &$^\circ$ \\ 
\noalign{\smallskip}\hline\noalign{\smallskip}
1	&18:03:20	&0.15 &20.5 &98 & 83.9 &+0.7 & 82 & 145--133 &34\\
2          &18:20:00	&2.3 &22.1 & 107& 93--90& $-$2.0& 83& 225--210&25\\
3	&19:58:22	&0.4 &35.6 & 102& 91--90, 88.3 &$-$0.4& 83.5 & 155--135&46 \\
4	&20:05:35	&3.3 & 36.5 & 107.5 & 96--93 & $-$2.3& 80.5 & 176--146&30 \\
5	&20:19:34	&0.4 & 38.7 & 101.5 & 89.6 & 0.0 & 84 & 125--104&2\\
6	&20:28:21	&2.7& 39.9 & 105 & 82.8 & $-$3.1 & 76.5 & 146--116&27  \\
7	&20:43:51	&0.3 & 41.3 & 101 & 96--93.5 & +0.3 & 86 & 185--170&15 \\
8	&20:55:34	&0.1 & 43.3 & 99 & 98--93.5 & +1.3 & 88.5 & 133--120&4 \\
\noalign{\smallskip}\hline
\end{tabular}
\end{table}

\begin{table}
\caption{Pre-atmospheric velocities, geocentric radiants, geocentric velocities, and orbits of the observed meteors 
(equinox J2000.0).}
\label{orbits}
\begin{tabular}{rrrrrrrrrrr}
\hline\noalign{\smallskip}
\# & 
\multicolumn{1}{c}{$v_\infty$} & 
\multicolumn{1}{c}{$\alpha_G$} & 
\multicolumn{1}{c}{$\delta_G$} & 
\multicolumn{1}{c}{$v_G$} & 
\multicolumn{1}{c}{$a$} & 
\multicolumn{1}{c}{$e$} & 
\multicolumn{1}{c}{$q$} & 
\multicolumn{1}{c}{$\omega$} & 
\multicolumn{1}{c}{$i$} & 
\multicolumn{1}{c}{$\Omega$} \\
&
\multicolumn{1}{c}{km/s} &
\multicolumn{1}{c}{$^\circ$} &
\multicolumn{1}{c}{$^\circ$} &
\multicolumn{1}{c}{km/s} &
\multicolumn{1}{c}{AU} &&
\multicolumn{1}{c}{AU} &
\multicolumn{1}{c}{$^\circ$} &
\multicolumn{1}{c}{$^\circ$} &
\multicolumn{1}{c}{$^\circ$}  \\
\noalign{\smallskip}\hline\noalign{\smallskip}
1&   23.44&  263.41&   55.49&   20.74&    3.48&   0.714&  0.9966&  173.62&   31.46& 194.949\\[-0.5ex]
 &   $\pm$.25&   $\pm$.29&   $\pm$.33&   $\pm$.28&   $\pm$.22&  $\pm$.018& $\pm$.0002&   $\pm$.23&   $\pm$.33\\
2&  23.55&  263.29&   55.52&   20.88&    3.56&   0.720&  0.9965&  173.54&   31.63& 194.960\\[-0.5ex]
 &   $\pm$.25&   $\pm$.33&   $\pm$.40&   $\pm$.28&   $\pm$.24&  $\pm$.019& $\pm$.0002&   $\pm$.26&   $\pm$.34\\
3&   23.20&  263.18&   55.66&   20.54&    3.28&   0.697&  0.9965&  173.48&   31.30& 195.027\\[-0.5ex]
 &   $\pm$.50&   $\pm$.25&   $\pm$.14&   $\pm$.57&   $\pm$.36&  $\pm$.033& $\pm$.0001&   $\pm$.21&   $\pm$.63\\
4&   23.57&  263.16&   55.74&   20.96&    3.54&   0.719&  0.9964&  173.50&   31.80& 195.032\\[-0.5ex]
 &   $\pm$.15&   $\pm$.14&   $\pm$.10&   $\pm$.17&   $\pm$.13&  $\pm$.010& $\pm$.0001&   $\pm$.12&   $\pm$.19\\
5&   23.48&  263.29&   55.96&   20.86&    3.42&   0.709&  0.9966&  173.65&   31.76& 195.042\\[-0.5ex]
 &   $\pm$.25&   $\pm$.28&   $\pm$.62&   $\pm$.28&   $\pm$.26&  $\pm$.022& $\pm$.0002&   $\pm$.24&   $\pm$.38\\
6&   23.55&  262.93&   55.75&   20.95&    3.51&   0.716&  0.9963&  173.33&   31.80& 195.048\\[-0.5ex]
 &   $\pm$.15&   $\pm$.10&   $\pm$.10&   $\pm$.17&   $\pm$.12&  $\pm$.010& $\pm$.0001&   $\pm$.10&   $\pm$.19\\
7&   23.38&  263.29&   55.52&   20.76&    3.47&   0.712&  0.9964&  173.51&   31.49& 195.058\\[-0.5ex]
 &   $\pm$.25&   $\pm$.19&   $\pm$.20&   $\pm$.28&   $\pm$.20&  $\pm$.017& $\pm$.0001&   $\pm$.15&   $\pm$.32\\
8&   23.63&  263.04&   56.04&   21.04&    3.51&   0.716&  0.9964&  173.49&   32.00& 195.066\\[-0.5ex]
 &   $\pm$.40&   $\pm$.53&    $\pm$3.8&   $\pm$.45&    $\pm$1.2&  $\pm$.094& $\pm$.0008&   $\pm$.80&    $\pm$1.4\\
\noalign{\smallskip}\hline
\end{tabular}
\end{table}

In Table~\ref{orbits},  the initial velocity at the entry into the atmosphere, $v_\infty$, the geocentric radiant, 
$\alpha_G$, $\delta_G$, geocentric velocity, $v_G$, and the orbital elements are given. The initial velocity was
determined from the modeling of deceleration by the erosion model \citep{Bor05}. The error of the velocity was 
estimated from the spread of the measurements.

\begin{figure}
  \includegraphics[width=1.0\linewidth]{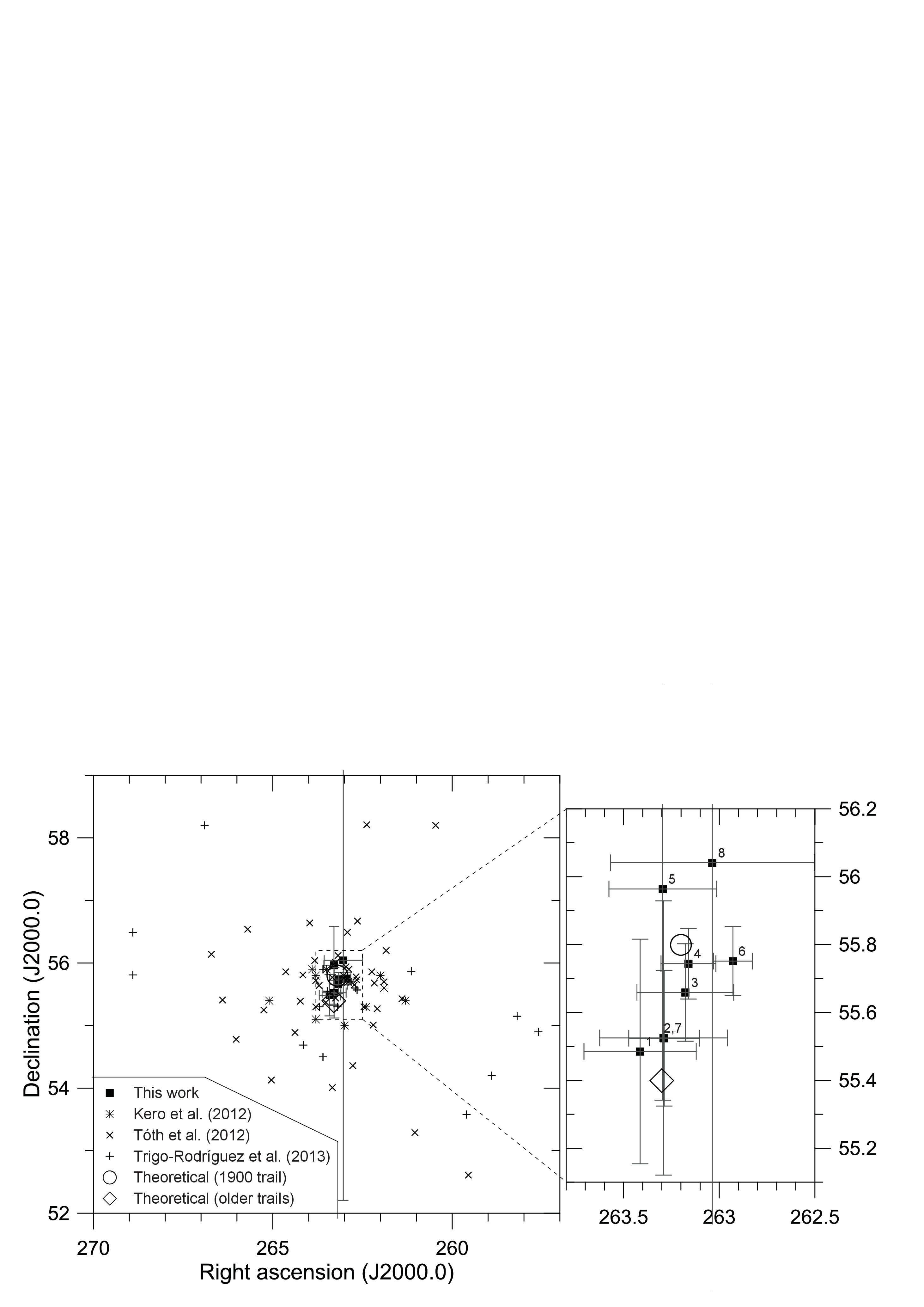}
\caption{Geocentric radiants of 2011 Draconid meteors as observed by \citet{Kero}, \citet{Toth}, \citet{Trigo},
and this work. Error bars are plotted for this work.
Theoretical radiants predicted by J.~Vaubaillon (http://draconids.seti.org) for the 1900
dust trail and 1873--1894 dust trails are given for comparison.
The inset on the right shows our data in detail and with meteors identified.}
\label{radiants}      
\end{figure}

The geocentric radiants are plotted and compared with the results of other authors in Fig.~\ref{radiants}.
Our radiants show lower scatter than those of other authors and are close to the 
theoretical predictions \citep{Maslov, Jeremy1}.
The meteors from the 1873--1894 trails were predicted to encounter the Earth earlier (about 17 UT) and to have radiants
$0.4^\circ$ to the South in comparison with the 1900 trail, which was responsible for the main activity peak around 20 UT. 
Judging from their time of appearance and radiant positions, it is possible that meteors 1 and 2 belonged to 
the older trails \citep[see also][]{IMC}. 
The most precise meteors 3, 4, and 6 almost certainly belonged to the 1900 trail.

\begin{figure}
  \includegraphics[width=1.0\linewidth]{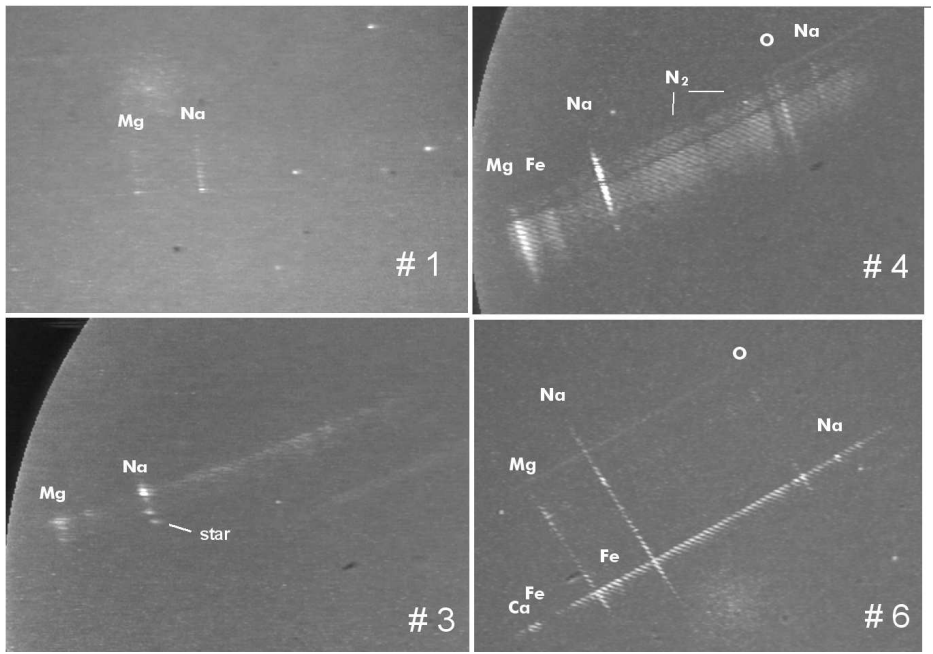}
\caption{Co-added video frames (with enhanced contrast) 
showing the spectra of meteors 1, 3, 4 and  6. In all cases the meteor flew from
upper left to bottom right and the wavelengths increase from bottom left to upper right. The main emissions are identified.
There is a small gap in the data of meteor 4 due to tape failure.}
\label{4spectra}     
\end{figure}

\begin{table}
\caption{Identification of spectral features in meteor 6.}
\label{identif}
\begin{tabular}{llllll}
\hline\noalign{\smallskip}
No. & \multicolumn{2}{c}{Observed} & \multicolumn{3}{c}{Identified} \\
& Wavelength & Intensity & Atom-Multiplet & Wavelengths & 2nd order  \\
\noalign{\smallskip}\hline\noalign{\smallskip}
1 & 3810w & 100 & Fe~{\sc I} - 20 & 3820, 3826, 3834 \\ 
&&& Mg~{\sc I} - 3 & 3838, 3832, 3829 \\
&&& Fe~{\sc I} - 4 & 3860, 3856, 3824 \\
&&& Fe~{\sc I} - 45 & 3816 \\
2 & 4250w & 35 & Fe~{\sc I} - 42 & 4326, 4308, 4272 \\
&&& Ca~{\sc I} - 2 & 4227 \\
&&& Cr~{\sc I} - 1 & 4254, 4275 \\
3 & 4380 & 55 & Fe~{\sc I} - 41 & 4384, 4405 \\
&&& Fe~{\sc I} - 2 & 4376 \\
4 & 4450 & 40  & Fe~{\sc I} - 2 & 4427, 4462, 4482 \\
5 & 4570 & 12 & Mg~{\sc I} - 1 & 4571 \\
6 & 4700 & 9 & Mg~{\sc I} - 11 & 4703 \\
7 & 4970w & 9 & Fe~{\sc I} - 318 & 4958, 4920 \\
8 & 5110 & 14 & Fe~{\sc I} - 1 & 5110 \\
9 & 5180 & 70 & Mg~{\sc I} - 2 & 5184, 5172, 5167 \\
&&& Fe~{\sc I} - 1 & 5166, 5169 \\
10 & 5260 & 18 & Fe~{\sc I} - 15 & 5270 \\
11 & 5330 & 12 & Fe~{\sc I} - 15 & 5328 \\
12 & 5420 & 14 & Fe~{\sc I} - 15 & 5405--5456 \\
13 & 5530 & 5 & Mg~{\sc I} - 9 & 5528 \\
14 & 5580 & 5 & Fe~{\sc I} - 686 & 5615, 5586, 5572 \\
&&& Ca~{\sc I} - 21 & 5588, 5594 \\
15 & 5890 & 100 & Na~{\sc I} - 1 & 5890, 5896 \\
16 & 6200--6800 && N$_2$ 1st. positive \\
17 & 7000--7500 && N$_2$ 1st. positive \\
18 & 7460 & 7 & N~{\sc I} - 3 & 7468, 7442, 7424 \\
19 & 7690 & 16 & K~{\sc I} - 1 & 7665, 7699 & Mg~{\sc I} - 3, Fe~{\sc I} - 4 \\
20 & 7770 & 60 & O~{\sc I} - 1 & 7772, 7774, 7775 \\
21 & 8210 & 40 &  N~{\sc I} - 2 &8185--8223  \\
&&&Na~{\sc I} - 4 &8195, 8183  \\
22 & 8470 & 20 & O~{\sc I} - 4 & 8446 & Ca~{\sc I} - 2 \\
\noalign{\smallskip}\hline
\end{tabular} \\
{\footnotesize
Wavelengths are given in \AA;, "w" means wide line. Intensities are in relative units, taking into account the
spectral sensitivity of the instrument. The continuum and molecular emissions were subtracted from the
intensities. Saturation of bright lines was taken into account.
Only the most important second order contributors are listed.}

\end{table}

\begin{figure}
  \includegraphics[width=0.9\linewidth]{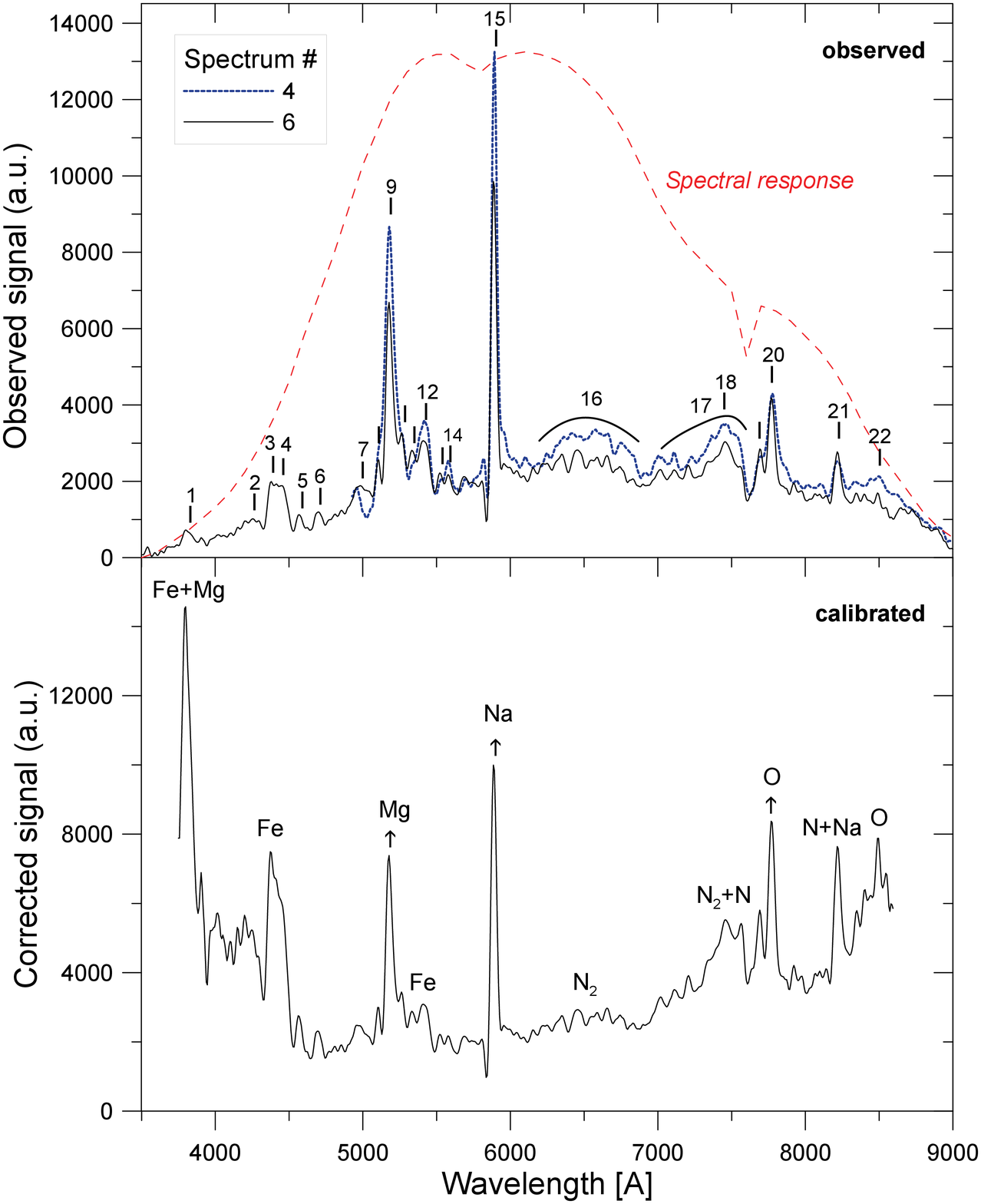}
\caption{Line identifications in Draconid meteor spectra. The upper panel shows the observed signal, 
integrated along the trajectory, as a function of wavelength for meteors 4 and 6. 
The blue part of spectrum 4 was outside the field of view. 
The spectral features identified in Table~\ref{identif} are marked. Note that the intensity of the
three strongest lines (9, 15, 20) is underestimated in the figure due to saturation during the meteor maximum.
The dashed line shows the relative spectral response of the system, measured using stellar spectra. The dip at 7600~\AA\
is due atmospheric O$_2$ absorption.
The lower panel shows spectrum 6 after correction to the spectral response. 
The intensity above 7500~\AA\ was affected by the admixture of second order spectrum. The minimum just left
of the Na line is an artifact.}
\label{calib}     
\end{figure}

\section{Spectra}
\label{spectrum}
The images of the four best spectra are reproduced in Fig.~\ref{4spectra}. The differences between meteor 
light curves shown in Fig.~\ref{LC}, 
are evident in these pictures. On the other hand, all spectra are similar with Mg and Na being the two brightest
lines. The plot of the two best spectra, integrated over the duration of the meteors, is given in Fig.~\ref{calib}.
The spectra contain significant continuous radiation over the whole wavelength range 3600--9000~\AA. 
The continuum provides about half of the signal. 
N$_2$ molecular bands are also present. The reliably identified emission features
are listed in Table~\ref{identif}. When corrected to the spectral sensitivity of the instrument, the blend of Fe and Mg 
lines near 3800~\AA\ becomes another strong feature. If compared with the Leonids
\citep{Leo98}, Draconids show lower intensities of the emissions of atmospheric origin (N, N$_2$, O), which
can be attributed to their much lower velocity in comparison to the Leonids. The lower velocity and lower intensity of
atmospheric lines also favors the visibility of the low excitation lines of Na and K in the infrared
(lines 19 and 21 in Table~\ref{identif}). They are seen in Draconids but not in Leonids. 
Otherwise, the meteoritic lines in Draconids are the same as seen in Leonids.

\begin{figure}
  \includegraphics[width=1.0\linewidth]{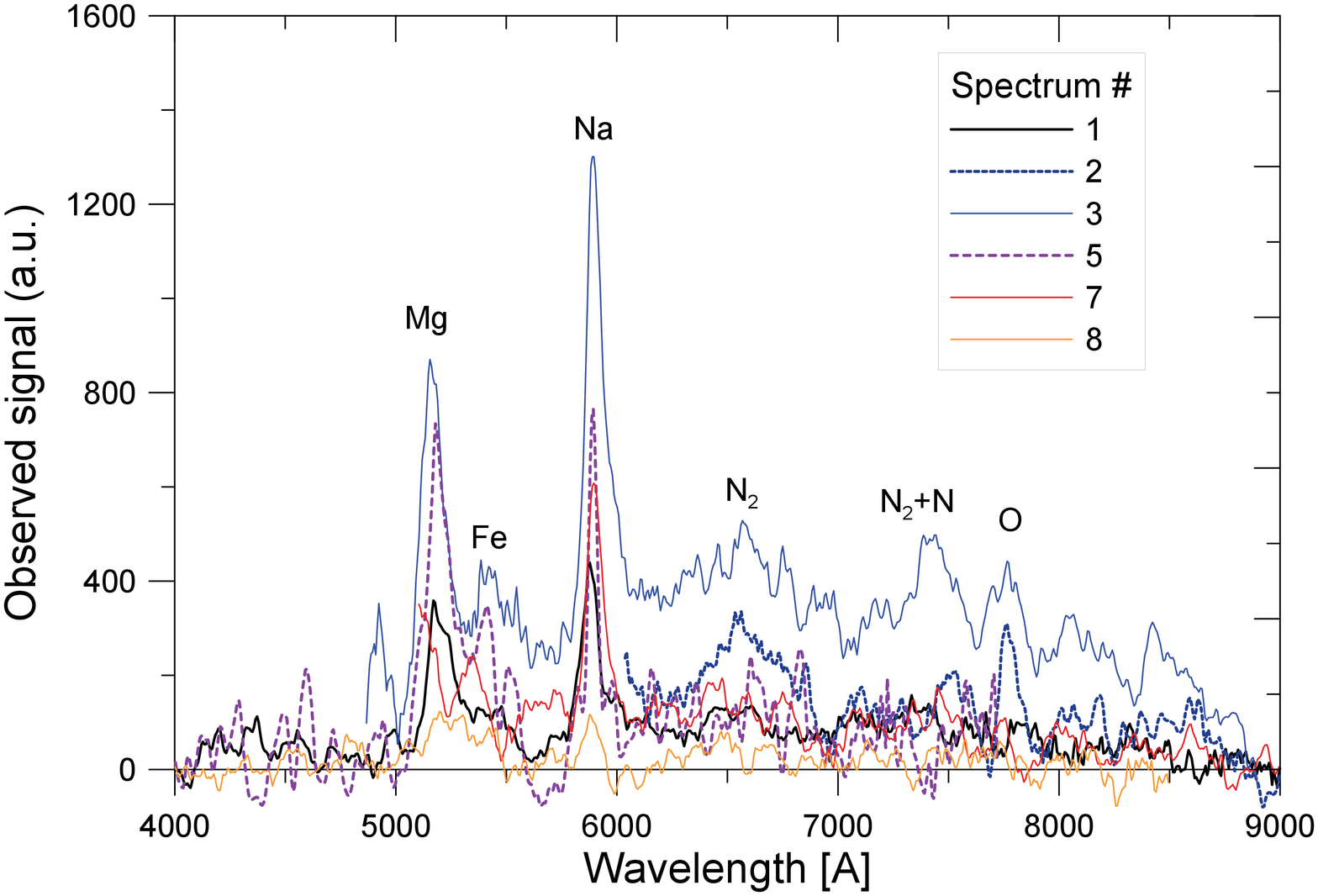}
\caption{Fainter Draconid spectra. Observed signal, integrated along the trajectory, is given as a function of wavelength.
The noise caused some curves to fall below zero after the background was subtracted.}
\label{totals}     
\end{figure}

\begin{figure}
  \includegraphics[width=0.9\linewidth]{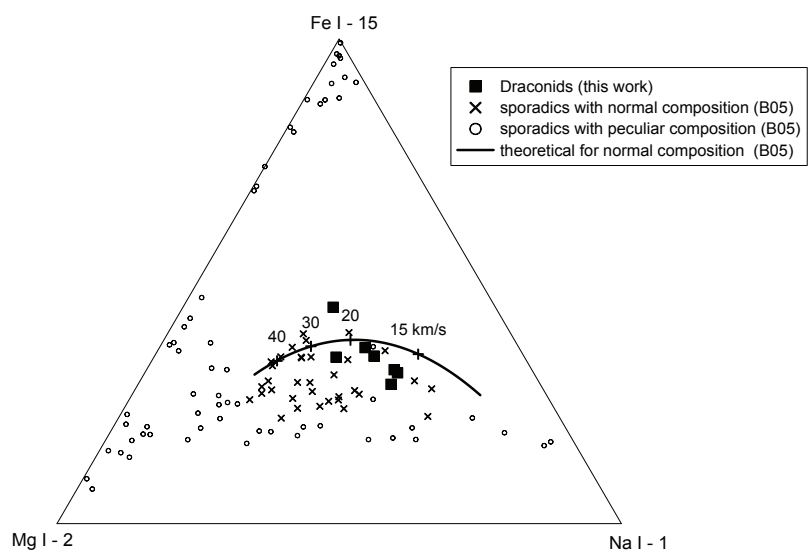}
\caption{Ternary diagram comparing the time-integrated intensities of Mg, Na, and Fe multiplets
in meteor spectra. Draconids studied in this work are plotted together with sporadic meteors studied by \citet[][=B05]{survey}.
Meteors classified by that work as having normal (i.e.\ chondritic) Mg-Fe-Na ratios are distinguished from meteors
with various peculiarities. A theoretical curve for thermal equilibrium
showing the expected positions of normal meteors as a function 
of their velocity is also shown \citep{survey}.}
\label{tridra}     
\end{figure}

The plots of the other six spectra are given in Fig.~\ref{totals}. These spectra have much lower
signal-to-noise ratio than spectra 4 and 6. Nevertheless, the intensities of Mg, Na, and Fe (multiplet 15) 
lines could be measured (except for meteor 2, where only the red part of the spectrum was captured). 
These lines have been used by \citet{survey} to compare the content of Mg, Fe, and Na in meteoroids
of various origin. As it can be seen in Fig.~\ref{tridra}, the Draconids fall in the region of meteors 
with normal, i.e.\ chondritic, 
composition. Other studies \citep{Millman, Bor05, Madiedo} also concluded that the
ratios of major elements are chondritic in Draconids. \citet{Madiedo} gave also the abundances 
of minor elements but their values are based on
low resolution and noisy spectra and cannot be considered as reliable.

\begin{figure}
  \includegraphics[width=0.9\linewidth]{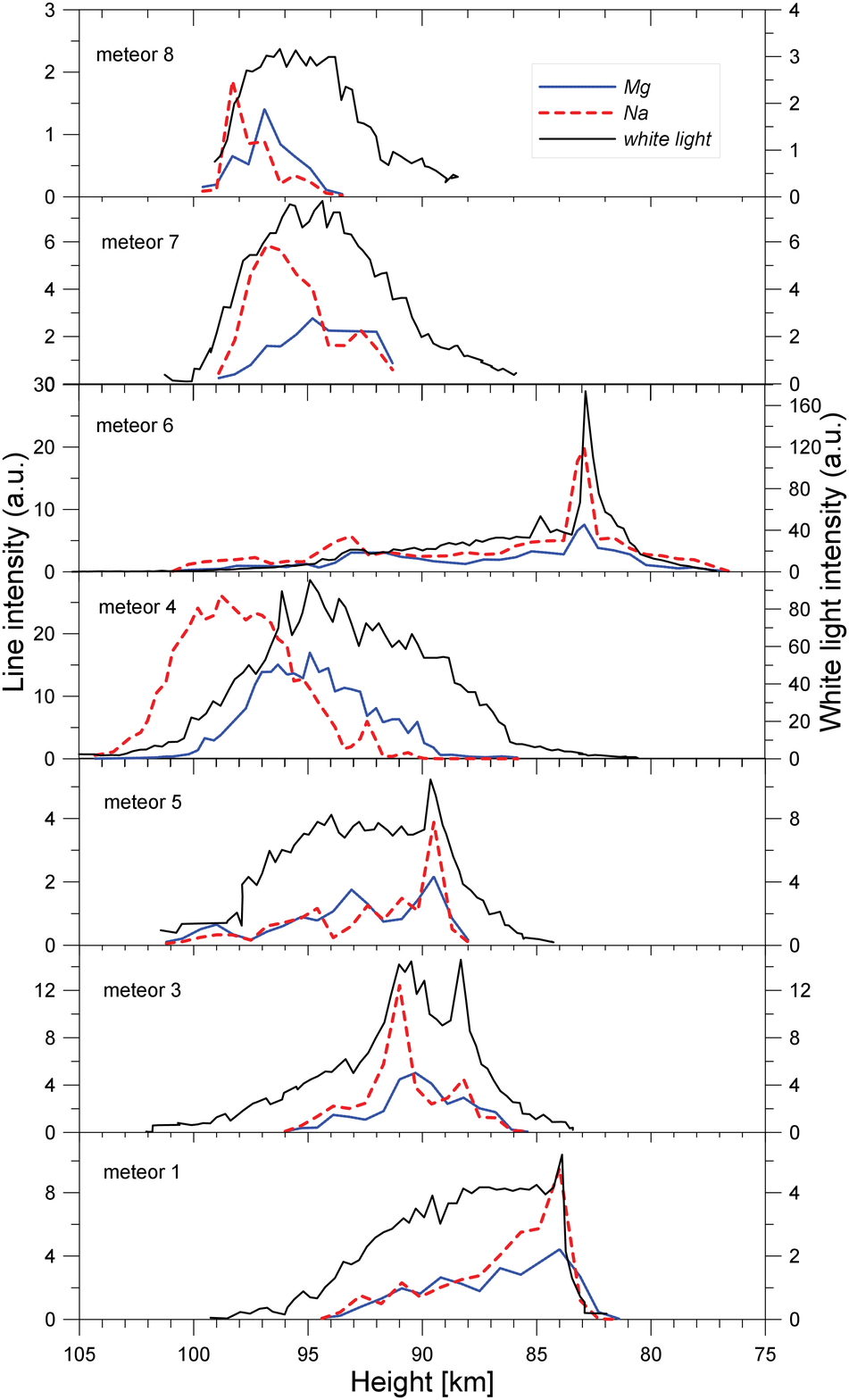}
\caption{Monochromatic light curves in the Mg and Na lines 
compared to the total light curves in white light. The intensities are in relative linear scale. The scale
in white light is different compared to the Mg and Na lines 
because the total intensity was measured on non-spectral cameras. The order
of the plots was adjusted to facilitate direct comparison of meteors 4 and 6.
Heights for the spectral curves are known to a precision of 1 km or better.}
\label{sodium}     
\end{figure}

Another aspect is the temporal evolution of the spectra. The previous studies \citep{Millman, Bor05}
noted a shift of the Na line toward higher altitudes in some Draconids. This was also the case of
a $-10$ mag fireball \citep{Madiedo}. 
Interestingly, our two best spectra 4 and 6 are quite different in this respect 
(Figs.~\ref{4spectra} and \ref{sodium}). Meteor 4 is a pronounced example of the early start and early end of
the sodium line. The maximum of Na is shifted up by about 5 km in comparison with Mg. In meteor 6, Na is present along
the whole trajectory in nearly constant proportion to Mg. Though both meteoroids had almost the same 
initial mass (Table~\ref{meteors}), the heights of the maxima and the shapes of the light curves were also quite different
(Figs.~\ref{LC} and \ref{sodium}). Meteor 4 had a flat maximum around height 95 km, while meteor 6 exhibited
a bright flare at a much lower height of 83 km. 
All these facts suggest that the meteoroid structure may be different.

\section{Deceleration}

In order to get more insight into the structure of the meteoroids, we evaluated their decelerations along the
trajectory. In fact, the directly measured quantity is the position of the meteor as a function of time (or frame number).  
Of course, individual position measurements are
subject to error. Moreover, once fragmentation starts, the meteor is no longer a point-like object.
As drag depends on mass, grains of different sizes separate along the meteor's direction of travel to form a streak of light
\citep{Bor05,Campbell}. As the fragmentation and ablation proceeds, different parts of the streak may become the brightest part.
This effect may lead to large scatter of apparent deceleration/acceleration from frame to frame. 
Thus, as a first step, we evaluated the lag of each meteor along its trajectory.
The lag at any  time is defined
as the difference between the predicted meteor position in case of zero deceleration (i.e.\ considering only
the initial position and velocity) and the actual position. Since the time is a relative quantity for each meteor, meteors
are best compared by the lag as a function of predicted height. The predicted height is the height as it would be without 
deceleration.

\begin{figure}
  \includegraphics[width=0.75\linewidth]{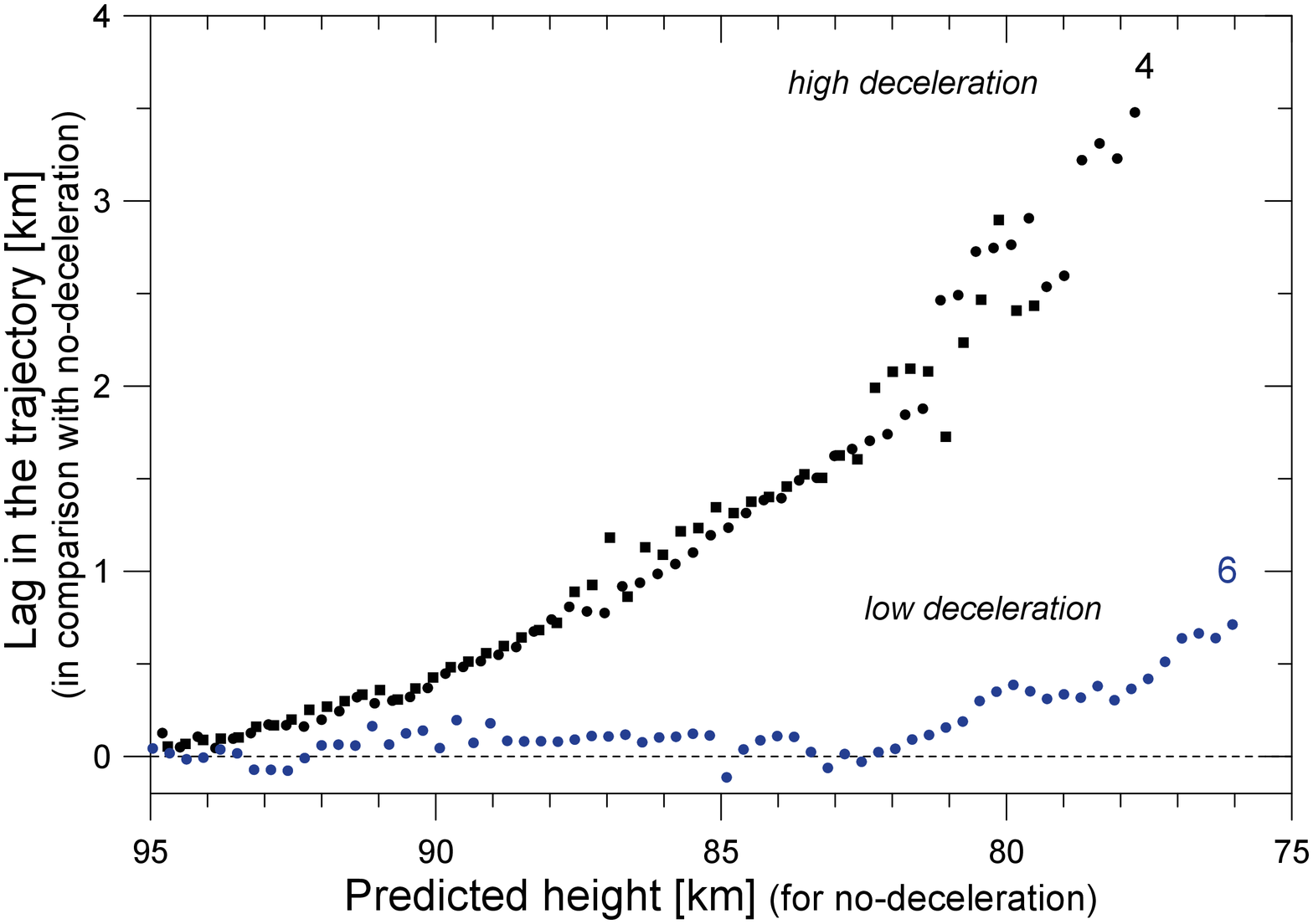}
\caption{Measured lag along the trajectory between the predicted position in case of no-deceleration and
the actual position as a function of predicted height for meteors 4 and 6. Different symbols for meteor 4 represent
the measurements from two different sites.}
\label{lag}
\end{figure}

\begin{figure}
  \includegraphics[width=0.75\linewidth]{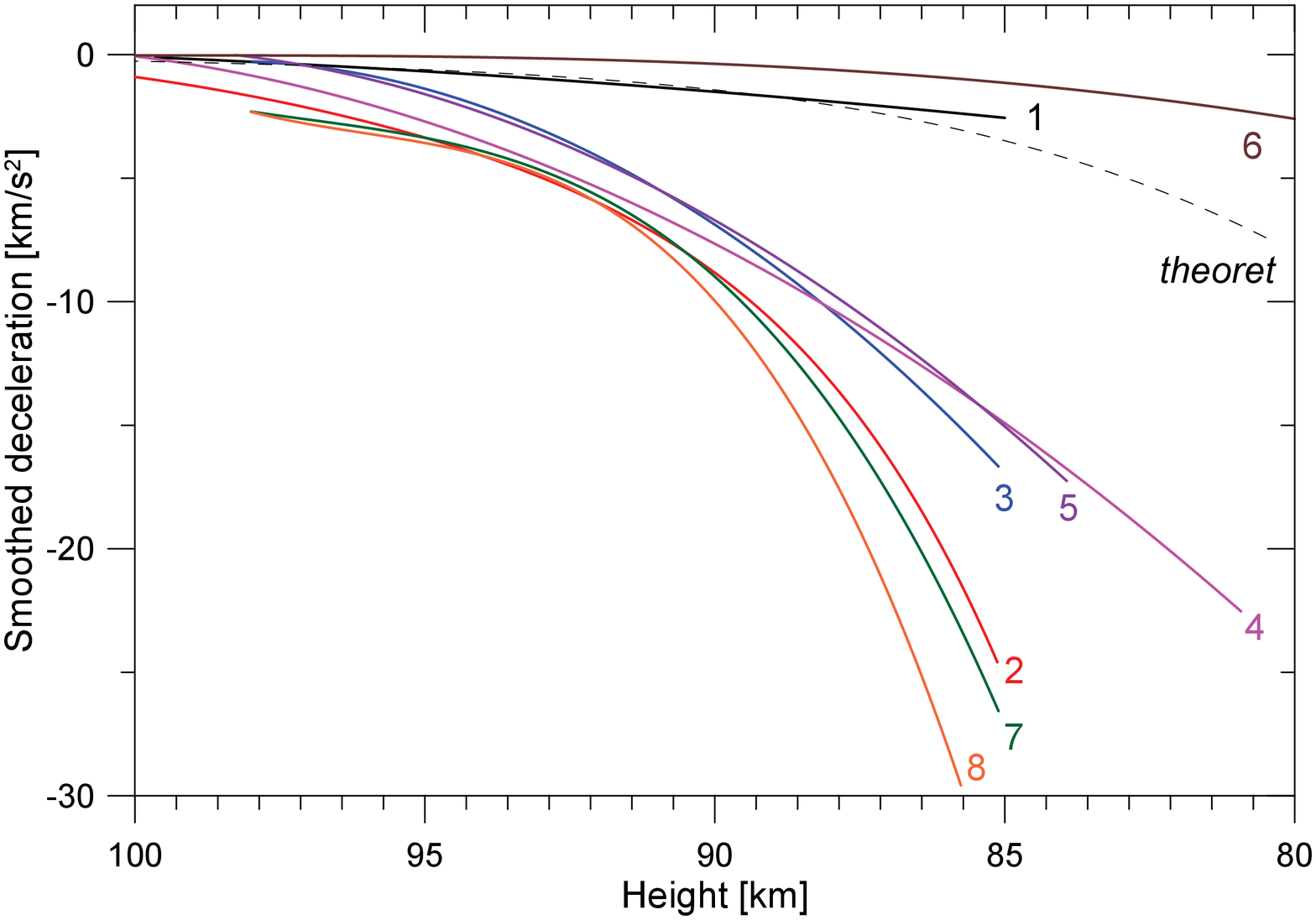}
\caption{Deceleration of eight Draconid meteors, smoothed 
by the erosion model, as a function of height. The dashed line labeled
\textit{theoret} shows the expected deceleration of a non-fragmenting spherical Draconid meteoroid of a mass of
$10^{-4}$ kg, bulk density 200 kg m$^{-3}$ (diameter 1 cm), ablation coefficient 0.02 s$^2$km$^{-2}$ and zenith angle
22$^\circ$. Meteor 6 showed less deceleration because it was more massive.}
\label{decel}
\end{figure}

The comparison of the lag of meteors 4 and 6 is given in Fig.~\ref{lag}. We can see that the lag (and thus the deceleration)
was much larger for meteor 4 than for meteor 6. This is another substantial difference between these two meteors,
besides the shape of the light curve and the release of sodium,
and again suggests differences in structure and fragmentation history.

The erosion model, described in detail by \citet{Bor05}, can provide further insight into the causes of the
differences between individual Draconid meteors. The model 
assumes that a gradual continuous release of individual grains, from which the meteoroid is composed,
starts at a certain height, after the meteoroid received a certain amount of energy per unit surface from the
collisions with atmospheric molecules. After the release, each grain behaves as an individual meteor. The grain density
is assumed to be 3000 kg m$^{-3}$. The bulk density of the whole meteoroids can be much lower due to high 
porosity. The rate of the grain release is described by the erosion coefficient, $\eta$. The rate of 
ablation (vaporization) of both the grains and the whole meteoroid is described by the ablation coefficient,
$\sigma$. The erosion may proceed in several (usually two) stages, i.e.\ the initial erosion may be applied only
to a certain percentage of the meteoroid, while the rest continues unaffected (only subject to ablation) for some
more time. The parameters of the erosion model are adjusted to fit the light curve in white light and the lag
in the trajectory.

To provide some
deceleration data, we smoothed the lag data fits obtained with the erosion model \citep{Bor05} and plotted 
the corresponding decelerations in Fig.~\ref{decel}. 
Only meteors 1 and 6 exhibited low deceleration.
We can see the decelerations of these two meteors were not higher than 
expected for non-fragmenting meteoroids of very low bulk density ($\sim 200$ kg m$^{-3}$) and similar mass.

\begin{table}
\caption{Selected parameters of the erosion model fits.}
\label{erosion}
\begin{tabular}{llllllllll}
\hline\noalign{\smallskip} 
\# & $E_{\rm S}$ & $f$ & $\eta$ & $\sigma$ & $h_{\rm es}$--$h_{\rm ee}$ & Sizes &
$E_{\rm S}$ (2) &  $\eta$ (2) & S (2) \\[0.5ex]
&MJ/m$^2$ && s$^2\!$/km$^{2}$ & s$^2\!$/km$^{2}$ & km & $\mu$m 
&MJ/m$^2$ & s$^2\!$/km$^{2}$  & $\mu$m\\
\noalign{\smallskip}\hline\noalign{\smallskip}
1 & 1.9 & 0.83 & 0.30 & 0.022 & 96--86 & 110& 16 & 0.8 & 5  \\
2 & 0.4 & 1.00 & 3.5 & 0.024 & 105--97 & 130 \\
3 & 1.5 & 0.50 & 0.6 & 0.027 & 98--94 & 170 & 4.2 & 0.7 & 80 \\
4 & 0.5 & 1.00 & 2.3 & 0.015 & 105--93 & 150--30\\
5 & 0.9 & 0.86 & 0.7 & 0.032 & 101--91 & 200--40 & 6.3 & 0.8 & 60 \\
6 & 1.1 & 0.84 & 0.15 & 0.015 & 101--78 & 130--20 & 23 & 1.0 & 5 \\
7 & 1.2 & 1.00 & 1.0 & 0.014 & 100--94  & 100--20  \\
8 & 1.4 & 1.00 & 1.0 & 0.019 & 100--95 & 100--20 \\
\noalign{\smallskip}\hline
\end{tabular}
{\footnotesize\flushleft\vspace{-1ex}
$E_{\rm S}$ is the energy received per unit cross-section before the erosion starts; $f$ is the fraction of mass of the
meteoroid subject to the initial erosion; $\eta$ and $\sigma$ are the erosion and ablation coefficients, respectively;
$h_{\rm es}$--$h_{\rm ee}$ is the height range of the initial erosion; Sizes is the size range of grains. Values
designated with (2) apply to the second stage erosion.

}
\end{table}

Table~\ref{erosion} gives the most important parameters of the erosion fits. The most fragile meteoroids 2 and 4
were characterized by an early start of erosion, high erosion rate, and the fact that their whole mass was subject to 
the initial erosion (although these meteoroids were among the largest  in our sample). 
Meteoroids 1 and 6 were characterized by much slower grain release, so that the erosion was finished
only at heights much lower than 90 km. Moreover, about 1/6 of both meteoroids did not participate in the initial 
erosion and resisted fragmentation to lower heights (84 and 83 km, respectively). Here they disrupted abruptly into
very small grains. On the other hand, 
there was no clear trend in grain sizes and ablation coefficients among the eight meteors.
We also computed the bulk densities but they are uncertain 
at least by a factor of two.
The obtained bulk densities were
between 100--200 kg m$^{-3}$ for meteors 2, 3, 7, and 8 and between 350--450 kg m$^{-3}$ for meteors 1, 4, 5, and 6. 
These differences may not be real; nevertheless, the erosion model confirms that Draconids are meteoroids with low bulk densities.

\begin{figure}
  \includegraphics[width=0.90\linewidth]{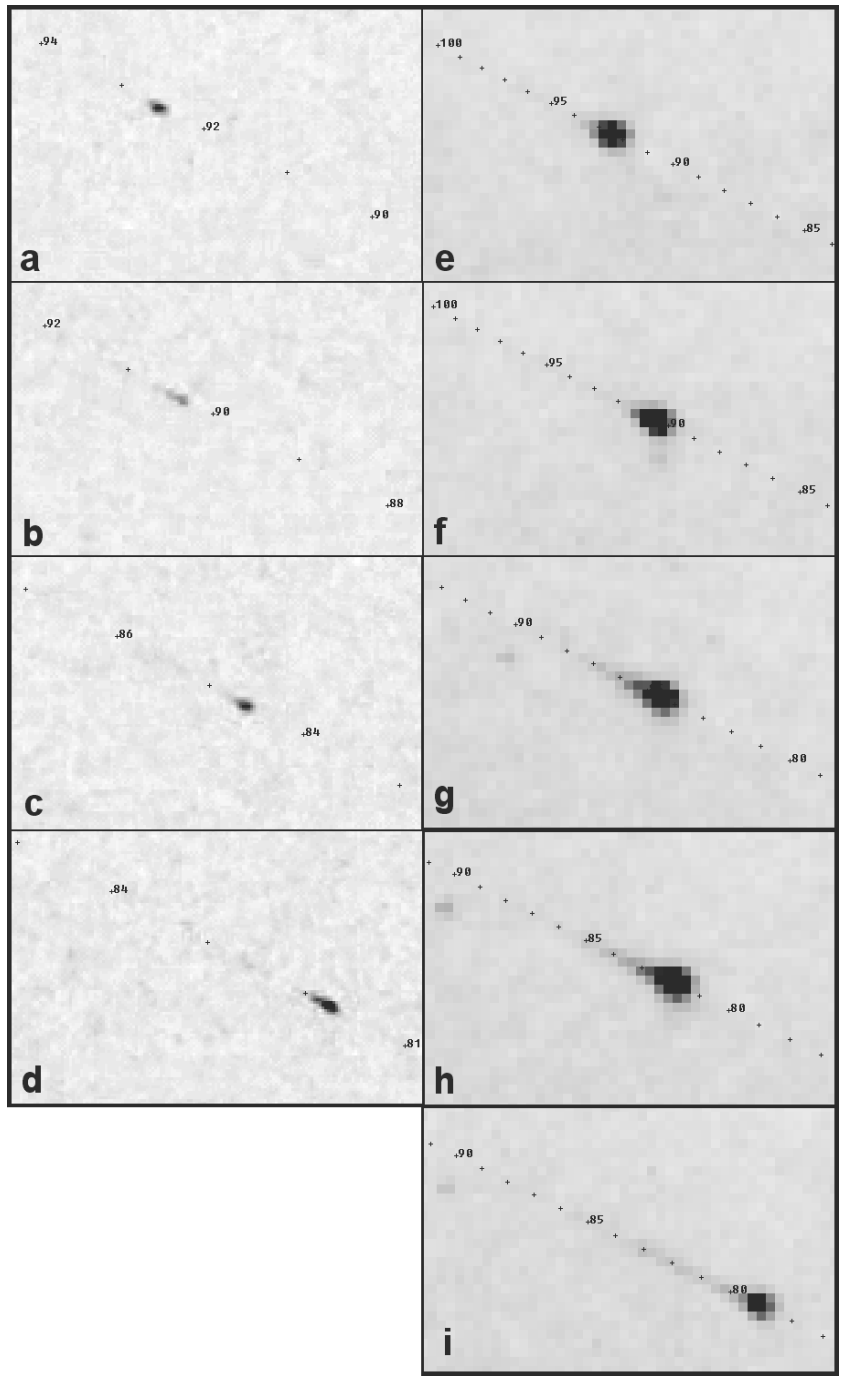}
\caption{Inverted images of meteor 6 at the heights 92.5, 90.5, 84.5, and 82 km from the HDV camera (panels a-d)
and the nearly corresponding images from the MAIA camera (panels e-h). The image from the MAIA camera at the
height 79 km is shown in panel i. The meteor was too faint for the HDV camera at that height.  Height marks
in the step of 1 km are given for the scale.}
\label{tvar6}      
\end{figure}

\begin{figure}
  \includegraphics[width=0.75\linewidth]{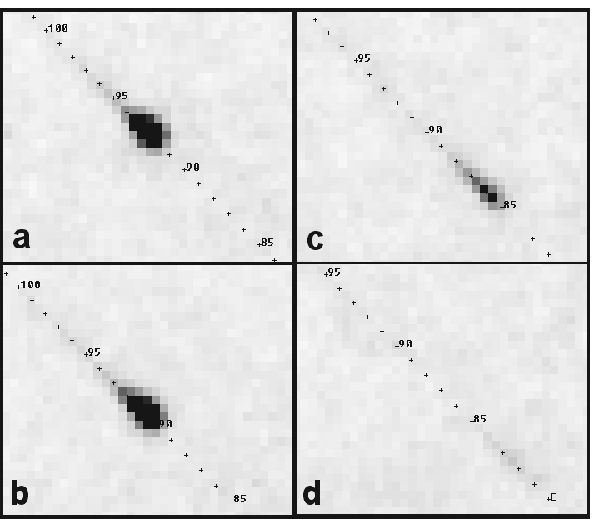}
\caption{Inverted images of meteor 4 at the heights 92, 90, 85, and 82 km from the MAIA camera. Height marks
in the step of 1 km are given for the scale.}
\label{tvar4}      
\end{figure}

\section{Meteor morphologies}

Besides light curves and decelerations, meteor morphologies, i.e.\ the shapes of meteor images, are indicative of the
fragmentation process \citep{Campbell}. With high resolution, individual fragments can be imaged or meteor wakes
caused by differential deceleration of grains can be measured. We got high resolution images of meteor 6 
with our non-intensified HDV camera. They are compared with the images from the low resolution but high sensitivity
MAIA camera in Fig.~\ref{tvar6}. Meteor height marks have been added to the images from both cameras.
The HDV camera provides the shape of the brightest part of the meteor, which was
too bright and unresolved on the MAIA images. 
For comparison, the images of meteor 4 from the MAIA camera are presented in Fig.~\ref{tvar4}. 

We can see that meteor 6 was still nearly point-like at the height of 92.5 km and only slightly elongated at 90.5 km. 
At 84.5 km, the meteor head was quite concentrated again but there was a faint wake seen in the MAIA camera,
about 2 km long. At 82 km, both the head and the wake became longer. At 79 km, the wake was more than 6 km long
but the meteor head was still well defined. Meteor 4 was clearly elongated in the low resolution MAIA image
already at the height of 92 km and the elongation increased with time. At lower heights, the meteor became a streak 
of light without any head. These differences are consistent with our interpretation that
meteoroid 4 quickly disintegrated into grains, which dispersed along the trajectory. Meteoroid 6 fragmented  more slowly
and in two stages. We note that the long wake may not have been caused only by the ablating grains, 
since Draconids also exhibit
persistent trains \citep{Trigo} whose radiation is driven by different mechanisms \citep{trains}. 

\section{Discussion}

We have combined spectroscopy, photometry, dynamics and morphology of Draconid meteors to study physical
and chemical properties of the meteoroids from comet 21P/Giacobini-Zinner. The meteoroids were somewhat larger
(sizes 1 -- 3 cm) than the majority of meteoroids from our previous Draconid study \citep{Bor05}. 
The spectra were not available in the previous study, except for one bright Draconid.

Meteors 2, 4, 7, 8 had some common characteristics: smooth and flat light curves without any flares, high
decelerations, and the release of most sodium in the first half of their trajectories (at least meteors 4, 7, 8 -- we do not
have complete spectrum of meteor 2). All these features can be explained by the complete and quick disintegration
of the meteoroids into small grains. In our model, the disintegration was finished at the height of 93 km or larger
for all four meteors. At the disintegration end height, 
almost all the sodium had evaporated. We believe that the reason was that the grains
were small enough to release sodium and potassium from their whole volumes earlier than the other elements
were vaporized, an effect called differential ablation \citep{Janches}. Alternatively, sodium may be part of 
a hypothetical glue which holds the grains together \citep{Hawkes}.

Meteors 1 and 6 exhibited flares at relatively low heights 84--83 km, where other Draconids had nearly disappeared.
The deceleration of meteoroids 1 and 6 was  much lower and sodium was present along the whole trajectories. 
Our analysis suggests
that their bulk density was not very different from the meteoroids of the previous group and their fragmentation
started at similar heights or only somewhat lower. The main difference was that either the grain release was 
much slower or they fragmented in a somewhat different manner. Moreover, significant parts of the meteoroids
(1/6 of mass in both cases) resisted the fragmentation until heights of 84--83 km. Here they disrupted abruptly causing
the meteor flares. These more compact
parts contained significant sodium and the sodium was radiated out efficiently during the flares (see Fig.~\ref{sodium}).
It is possible that mechanical strength of the material was exceeded at these heights, where
the dynamic pressure reached 5 kPa. This would still mean  a quite low mechanical strength in comparison with
most non-Draconid meteoroids \citep{IAUS229}. 

Meteoroids 3 and 5 were intermediate cases. 
The light curves and the sodium release pattern were similar to meteors 1 and 6
but the decelerations were higher and the flares occurred at larger altitudes. Meteoroid 3 was particularly complex
since it exhibited three stages of erosion/fragmentation.

We note that meteors 1 and 2, which may have come from an older ejection event than the others, 
have very different characteristics, and are each more similar to meteors from the main peak than to each other. 
This fact suggests that the differences found among the meteoroids reflect the inhomogeneity of the cometary material
and not the exposure age in space.

\section{Conclusions}

We analyzed eight moderately bright Draconid meteors. Some of the most precise trajectories and orbits of the
2011 Draconids were provided. The main purpose of this work was, nevertheless, to study the physical and chemical 
properties of Draconid meteoroids.  Our data are consistent with the known fact that Draconids are porous
conglomerates of grains and the abundance ratios of the main elements (Mg, Fe, Na) are chondritic. 
Nevertheless, significant differences in fragmentation behavior of cm-sized Draconids were found. Various textures
with various resistance to atmospheric fragmentation clearly exist among Draconid meteoroids and even 
within single meteoroids.

\begin{acknowledgements}
We thank the other expedition participants, Jaroslav Bo\v cek and Vlastimil Voj\'a\v cek, for their
assistance. Pavel Spurn\'y navigated us remotely to the clear sky region. Karel Fliegel 
and Stanislav V\'\i tek prepared the MAIA cameras. 
We thank the referees, Margaret Campbell-Brown, 
Edward Stokan, and an anonymous referee,
for their comments and improvements of the text. 
This work was supported
by grants P209/11/1382 and P209/11/P651 from GA\,\v CR. System MAIA was developed within 
the GA\,\v CR grant 205/09/1302. The institutional project was RVO:67985815.
\end{acknowledgements}


\end{document}